\documentclass[aps,twocolumn,showpacs,pra,amsmath,amssymb,floatfix,superscriptaddress]{revtex4-2}
\usepackage{amsmath,amssymb,graphicx,color}
\usepackage{bm}
\usepackage{bbm}
\usepackage[caption=false]{subfig}
\usepackage[usenames,dvipsnames]{xcolor}
\usepackage[normalem]{ulem}
\usepackage{soul}
\usepackage[colorlinks=true,citecolor=blue,linkcolor=RubineRed,urlcolor=blue]{hyperref} 
\usepackage{soul,xcolor}
\usepackage{cases}

\usepackage{amsmath}             
\usepackage{amssymb}             
\usepackage{amsfonts}            
\usepackage{mathrsfs}            

\usepackage{physics}

\usepackage{mathptmx} 

\DeclareMathAlphabet{\mathcal}{OMS}{cmsy}{m}{n}
\DeclareSymbolFont{largesymbols}{OMX}{cmex}{m}{n}

\usepackage{empheq}

\begin{document}
	\title{Persistent current by a static non-Hermitian ratchet}
	
	\author{Guitao Lyu}\email{guitao@zju.edu.cn}
	\affiliation{Department of Physics and Zhejiang Institute of Modern Physics, Zhejiang University, Hangzhou, Zhejiang 310027, China}	
	\author{Gentaro Watanabe}\email{gentaro@zju.edu.cn}
	\affiliation{Department of Physics and Zhejiang Institute of Modern Physics, Zhejiang University, Hangzhou, Zhejiang 310027, China}
	\affiliation{Zhejiang Province Key Laboratory of Quantum Technology and Device, Zhejiang University, Hangzhou, Zhejiang 310027, China}
	\date{\today}
	
	\begin{abstract}		
          We propose a scheme to generate a persistent current in driven-dissipative systems which can be described by the generalized Gross-Pitaevskii (GP) equation. Our proposal consists of fabricating a rachet-potential shape of the loss-rate profile, which simultaneously breaks the time-reversal symmetry and the parity-inversion symmetry. Unlike existing schemes to generate a current using a rachet potential in Hermitian systems, no dynamic drive is needed. The basic physics of our scheme is discussed by a simple discrete driven-dissipative GP model, and the results are also verified by a realistic continuous model. Furthermore, we demonstrate the experimental feasibility of our scheme to generate the persistent current in exciton-polariton condensates in a semiconductor microcavity.
	\end{abstract}
	
	\maketitle

\section{Introduction}

Persistent current of particles in superfluids and superconductors is a paradigm of macroscopic quantum phenomena. The phenomenon is not only interesting by itself from the perspective of fundamental physics but also it has many applications. It is useful for constructing quantum interference devices \cite{Simmonds2001nature, Hoskinson2006PRB, Wang2005PRL, Sturm2014ncomms} thanks to its matter-wave properties. Systems that can generate a current are also used for some information processing devices \cite{Seaman2007PRA, Pepino2009PRL, Amo2010natphoton, Buluta2011RPP, Gao2012PRB, Piccione2012natnanotech, Ballarini2013natcommun, Wendin2017RPP, Zasedatelev2019natphoto, Blais2020natphys}. Another interesting application of generating a current can be found in atomtronic circuits on an atomic chip \cite{Dekker2000PRL, Folman2000PRL, Nirrengarten2006PRL, Mukai2007PRL, Wang2005PRL, Seaman2007PRA}, which is an analog of an electronic chip using the current of cold atoms.

Using a time-periodically driven ratchet potential \cite{Reimann1997PRL, Flach2000PRL, Reimann2002PhysRep, Astumian2002PhysToday, Denisov2014PhysRep} is a typical scheme to generate a current which is widely employed in various systems such as cold atoms \cite{Poletti2007PRA, Creffield2009PRL, Heimsoth2010PRA, Denisov2014PhysRep}, Brownian particles \cite{Astumian2002PhysToday}, superconducting circuits with a Josephson junction array \cite{Zapata1996PRL, Goldobin2001PRE, Sterck2002APA, Falo2002APA, Majer2003PRL, Shalom2005PRL}, etc. Even if a driving force is unbiased on average, a nonzero net current can be generated because the time-reversal ($\mathbb{T}$) and the parity-inversion ($\mathbb{P}$) symmetries of the system are simultaneously broken by the time-dependent ratchet potential (see, e.g., Refs.~\cite{Reimann2002PhysRep, Denisov2014PhysRep}). See also Refs.~\cite{Flach2002PRL, Molina2008NJP} for generating a directed current in nonlinear systems using a drive without shift-symmetry. Another conventional way to generate a current is using a gauge potential, for instance, to stir cold atoms \cite{Fetter2009RMP, Babiker2019JO} or exciton-polariton condensates \cite{Ramanathan2011PRL, Kwon2019PRL} by a laser with orbital angular momentum. A current of overdamped particles with complex polarizability can also be generated by a vector potential  \cite{Zapata2009PRL}. Recently, several schemes to generate a persistent current using nonlocal dissipation have also been discussed  \cite{Metelmann2015PRX, Keck2018PRA, Yamamoto2020PRR}.

In this paper, we propose a scheme to generate a persistent current in driven-dissipative systems described by a non-Hermitian Hamiltonian. The exciton-polariton condensate system in a semiconductor microcavity \cite{Keeling2007review, Deng2010RMP, Carusotto2013RMP, Byrnes2014natphys, Deveaud2015AnnuRev} is a typical example of the platform for this proposal. An exciton-polariton, a quasiparticle consisting of a semiconductor exciton coupled with a microcavity photon, can form Bose-Einstein condensates (BECs) at high temperatures, even at room temperature \cite{Plumhof2014natmater, Sedov2014PRX}, due to its very small effective mass. Although the lifetime of an exciton-polariton is finite ($O(10^2)\, \text{ps}$ \cite{Nelsen2013PRX, Sedov2014PRX}) due to the leakage of the microcavity photons, the condensates can be sustained when the loss is compensated by the external laser pumping. Such driven-dissipative systems show many intriguing nonequilibrium phenomena. Generation of a persistent current in exciton-polariton condensates has come into view recently \cite{Saito2016PRB, Sedov2021PRR, Lukoshkin2018PRB, Gallemi2018PRB, Stroev2020PRB}.

Our proposal is to fabricate the loss-rate profile in the form of the ratchet potential, which breaks the $\mathbb{P}$ symmetry of the system. Since the loss and gain break the $\mathbb{T}$ symmetry, such a ratchet-potential-shaped loss-rate profile, even if it is time independent, simultaneously breaks the $\mathbb{T}$- and $\mathbb{P}$symmetry of the system, so that it can generate a net current. Unlike the Hermitian Hamiltonian with a ratchet potential, which is required to drive the potential in a time-dependent manner to break the $\mathbb{T}$ symmetry of the system, such a dynamical drive is unnecessary in our scheme. In addition, our scheme can generate a nonzero net current even in the thermodynamic limit. This is in contrast to the scheme using a gauge potential in Hermitian systems and the one using an elliptic pump together with a finite cylindrical pillar microcavity for the exciton-polariton condensates \cite{Lukoshkin2018PRB, Sedov2021PRR} whose current vanishes in the limit of the infinite radius of the system. Moreover, the minimum width of the flow by these schemes is bounded by the size of the laser spot, which is in the order of micrometers, while that of our scheme can go down to the nanoscale by the semiconductor processing technology \cite{Das2011PRL, Piccione2012natnanotech}. Compared to the nonlocal dissipation proposed in Refs.~\cite{Metelmann2015PRX, Keck2018PRA, Yamamoto2020PRR} which is not easy to implement in experiments, our scheme employs only the local dissipation with the position-dependent loss rate, which can be realized by tuning the quality factor of the microcavity by etching the distributed Bragg mirrors \cite{Dousse2008PRL, Tanese2013natcommun, Jayaprakash2020ACSPhoto}.

In this paper, first we employ a simple model based on the discrete driven-dissipative Gross-Pitaevskii (GP) Hamiltonian to discuss the basic physics. The effects of the violation of the $\mathbb{P}$ symmetry of the initial state and that of the effective pump and loss on the current generation are discussed. The effects of the gain-saturation and the system size on the steady current are also clarified. Then, we turn to a more realistic continuous model (the generalized Gross-Pitaevskii equation \cite{Keeling2008PRL, Eastham2008PRB, Wouters2010PRL, Keeling2011Contempphys, Moxley2016PRA}) to verify the results obtained by the discrete model. Finally, we demonstrate the generation of a persistent current of the exciton-polariton condensates in a semiconductor microcavity, by the widely used coupled GP equations \cite{Wouters2007PRL, Carusotto2013RMP, Ma2015PRB, Chestnov2016PRB}.

\section{Discrete model}\label{sec:discrete model}

We consider a one-dimensional (1D) time-independent driven-dissipative Bose system with the spatially periodic external potential $V$, the pump $P$, and the loss $\gamma$ with period $d$. To get a clear understanding of the basic physics, first we employ a simple model based on the discrete driven-dissipative GP Hamiltonian, which provides a simplified description of the system by sampling only a few points (specifically, four points in our model) per period $d$ of the external parameters (i.e., $V$, $P$, and $\gamma$). Each of the points is called a ``site" and each period $d$ with four sites is a unit ``cell." The Hamiltonian is given by
\begin{equation}
\begin{aligned}
H =&-J \sum\limits_{j} (\psi_{j}^{*} \psi_{j+1} +\psi_{j+1}^{*} \psi_{j}) + \dfrac{U}{2} \sum\limits_{j} |\psi_{j}|^4\\ 
&+ \sum\limits_{j}V_{j}|\psi_{j}|^2 +\sum\limits_{j} i(P_{j} - \gamma_{j})|\psi_{j}|^2 ,
\end{aligned}
\label{eq:1}
\end{equation}
where $j$ labels each site of the system, $\psi_{j}$ is the amplitude of the wave function at site $j$ with the normalization $\sum\limits_{j}|\psi_{j}|^2$ being the total number of particles, $ J $ represents the hopping matrix element between the neighboring sites, and $ U $ is the on-site interaction parameter. Here, the interaction parameter $U$ is complex: its real part describes the interaction between the particles at the same site, and its imaginary part describes the gain-saturation effect \cite{Keeling2008PRL, Keeling2011Contempphys}. The gain-saturation term, which is commonly employed in the effective GP model for exciton-polariton condensates, is usually needed to have a steady state. Without this term, the total number of particles will diverge or decay to zero once the net effects of the pump $ P_{j} $ and the loss $\gamma_{j}$ are imbalanced. For convenience, we introduce the effective pump $ F_{j} $ defined as $ F_{j} \equiv P_{j} - \gamma_{j} $, which can be regarded as an imaginary external potential. Since the imaginary potential $F_{j}$ breaks the time-reversal ($\mathbb{T}$) symmetry, we can simultaneously break the parity-inversion ($\mathbb{P}$) symmetry and the $\mathbb{T}$ symmetry without the time-dependent drive by introducing the ratchet-potential form of $ F_{j} $. An arbitrary function $f$ of $x$ has the $\mathbb{P}$ symmetry if and only if there exists some appropriate spatial shift $\Delta x$ which satisfies the relation $f(x+\Delta x)= f(-x)$. For the discrete model, this relation reads $f_{j+\Delta j}=f_{-j}$, where $f_{j}$ represents $f$ at site $j$. Here, it is noted that the spatial shift $\Delta j$ does not have to be an integer. 

The equation of motion of this system is given by
\begin{align}
i \hbar \dfrac{d\psi_{j}}{dt} &= \dfrac{\partial H}{\partial\psi_{j}^{*}} \notag\\
                   &= -J (\psi_{j+1} + \psi_{j-1}) + U|\psi_{j}|^2\psi_{j} + V_{j}\psi_{j} + i F_{j}\psi_{j}.
\label{eq:2}
\end{align}
This is the generalized discrete nonlinear Schr\"odinger equation (gDNLSE). The average current of particles per cell is defined as
\begin{equation}
\begin{aligned}
\mathcal{J} = \frac{i J}{2 N_{\text{cell}} \hbar } \sum\limits_{j} (\psi_{j} \psi_{j+1}^{*}  -  \psi_{j}^{*} \psi_{j+1} ),
\end{aligned}  
\label{eq:3} 
\end{equation}  
where $N_{\text{cell}}$ is the number of cells of the system. We take $J$ as the unit of energy and $\hbar/J $ as the unit of time in the following analysis of the discrete model. 

Now, we study the effect of violation of the $\mathbb{P}$ symmetry of the initial state $\psi_{j}^{(0)} $ and that of the external potential for both the real ($ V_{j} $) and the imaginary ($ F_{j} $) ones. For simplicity, we consider only a single cell which contains four sites and the wave function satisfies the periodic boundary condition. For now, we set $ U=0 $ and $ F_{j} $ is tuned to balance between the net effects of the pump $ P_{j}  $ and the loss $ \gamma_{j} $ (this fine-tuning is not necessary when we take a nonzero value for the imaginary $U$ in the later discussion). In our numerical calculation, for a $\mathbb{P}$-symmetric and a $\mathbb{P}$-asymmetric $F_j$, we take

\begin{empheq}[left={F_j=\empheqlbrace}]{alignat=2}
	&F_0 \sin{\left[\frac{\pi}{2} (j+2)\right]} & \quad \mbox{($\mathbb{P}$-symmetric)}\,,\medskip\label{eq:fjsym}\\
	&F_0 \left[ a - \text{mod}(j+2, 4)\right] & \quad \mbox{($\mathbb{P}$-asymmetric)}\,,\label{eq:fjasym}
\end{empheq}

respectively, with the amplitude factor $F_0=J/2$ for both cases and the tuning offset $a$ between the pump and the loss being $a=1$. For the real potential $V_j$, we use the same functions as Eqs.~(\ref{eq:fjsym}) and (\ref{eq:fjasym}) (with the same values of $F_0=J/2$ and $a=1$) for the $\mathbb{P}$-symmetric and the $\mathbb{P}$-asymmetric cases, respectively.

\begin{figure}[tb!] 
	\centering	
	\includegraphics[width=1\linewidth]{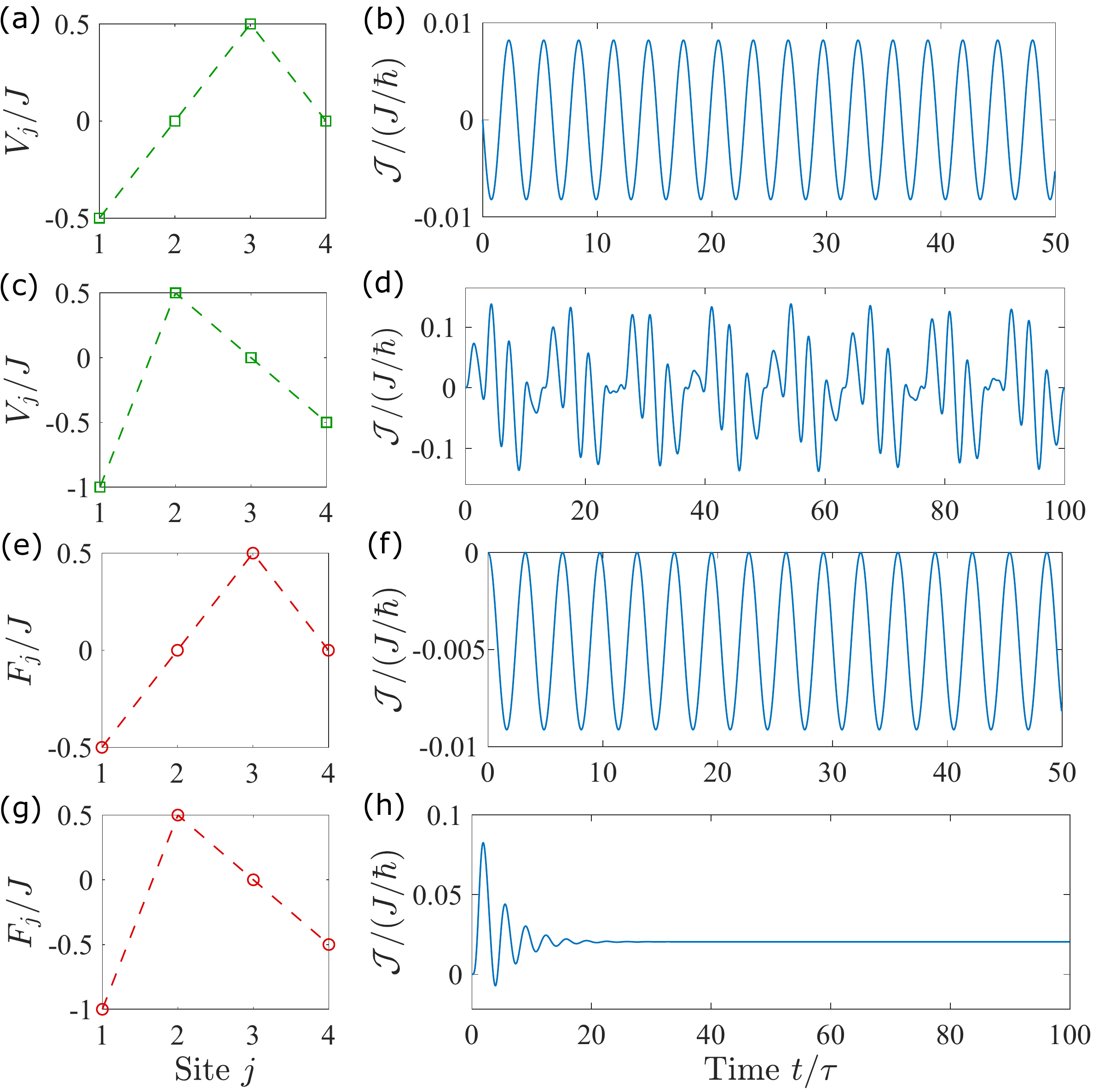} 
	\caption{$\mathbb{P}$ symmetry of the external potential and the resulting current. The real ($V_{j}$) and the imaginary ($F_{j}$) potentials are shown by the green and red dashed lines in the left panels, respectively. The right panels show the current as a function of time $t$ for the cases corresponding to their respective left panels. The first row [panels (a) and (b)]: $V_{1 \sim 4}/J = \{-0.5,\, 0,\, 0.5,\, 0 \}$ is $\mathbb{P}$ symmetric, $ F_{j} = 0 $, and $\psi_{j}^{(0)}$ (with random perturbation) is $\mathbb{P}$ asymmetric. The second row [panels (c) and (d)]: $V_{1 \sim 4}/J = \{-1,\, 0.5,\, 0,\, -0.5 \}$ is $\mathbb{P}$ asymmetric, $ F_{j} = 0 $, and $\psi_{j}^{(0)}$ (without random perturbation) is $\mathbb{P}$ symmetric. The third row [panels (e) and (f)]: $ V_{j} = 0 $, $F_{1 \sim 4}/J = \{-0.5,\, 0,\, 0.5,\, 0 \}$ is $\mathbb{P}$ symmetric, and $\psi_{j}^{(0)}$ (with random perturbation) is $\mathbb{P}$ asymmetric. The bottom row [panels (g) and (h)]: $ V_{j} = 0 $, $F_{1 \sim 4}/J = \{-1,\, 0.5,\, 0,\, -0.5 \}$ is $\mathbb{P}$-asymmetric, and $\psi_{j}^{(0)}$ (without random perturbation) is $\mathbb{P}$ symmetric. Here, we set $U=0$.}
	\label{fig:1}
\end{figure}

In the case of a static real potential ($ V_{j} \neq 0 $ and $F_{j} =0 $), there is no current at any time if $ V_{j} $ and $ \psi_{j}^{(0)} $ are $\mathbb{P}$ symmetric. If the $\mathbb{P}$ symmetry of the initial state $\psi_{j}^{(0)} $ is broken, the system has a nonzero current even if $ V_{j} $ is $\mathbb{P}$ symmetric, but the long-time averaged net current is still zero. As an example, we consider the following initial state whose particle number per cell is normalized to unity:
\begin{align}
\psi_{j}^{(0)}=\frac{ \displaystyle{\sqrt{N_{\text{cell}}} \left( 1+\xi_{j} \right) } } { \sqrt{ \sum_{j=1}^{N_{\text{site}}}  \left| 1+\xi_{j} \right|^{2} } },
\label{eq:4}
\end{align}
where $ N_{\text{site}} $ is the total number of sites of the whole system and $ \xi_{j} $ is the random perturbation. We take the random perturbation with a uniform distribution in the range of $\xi_{j}\in [-0.05, 0.05]$ for the initial state $ \psi_{j}^{(0)} $. Here, we take a $\mathbb{P}$-symmetric $ V_{j} $ as shown by the green dashed line in Fig.~\ref{fig:1}(a). Due to the random perturbation added, the initial state $ \psi_{j}^{(0)} $ is no longer $\mathbb{P}$ symmetric. The $\mathbb{P}$-symmetry breaking of $ \psi_{j}^{(0)} $ results in a nonzero current as shown in Fig.~\ref{fig:1}(b). However, even though the instantaneous current takes nonzero values, the average current oscillates around zero periodically in time and its long-time averaged value is zero. On the other hand, once the $\mathbb{P}$ symmetry of $ V_{j} $ is broken (regardless of whether $\psi_{j}^{(0)}$ is with or without random perturbation), e.g., by employing a ratchet-potential form of $V_{j}$ as shown in Fig.~\ref{fig:1}(c), the current $\mathcal{J}$ oscillates around zero irregularly [see Fig.~\ref{fig:1}(d)] (a long-time averaged value of the current is still zero though).

In the case of the static imaginary potential ($V_{j} =0 $ and $ F_{j} \neq 0 $), similarly the current is zero at any time if both $ F_{j} $ and $\psi_{j}^{(0)}$ have the $\mathbb{P}$ symmetry. However, if there is random perturbation $\xi_{j}$ in $\psi_{j}^{(0)}$, which breaks the $\mathbb{P}$ symmetry of $\psi_{j}^{(0)}$, a nonzero net current $\mathcal{J}$ is generated. Here, the current $\mathcal{J}$ periodically oscillates on either the positive or the negative side depending on the initial random perturbation [see Fig.~\ref{fig:1}(f) showing a case in which the current oscillates on the negative side]. Once the $\mathbb{P}$ symmetry of $ F_{j} $ is broken, e.g., by taking a ratchet-potential form as shown by the red dashed line in Fig.~\ref{fig:1}(g), the system has a nonzero steady current [see Fig.~\ref{fig:1}(h)] provided the net effects of $ P_{j} $ and $ \gamma_{j} $ are balanced (otherwise the current will diverge or decay to zero).

We now provide a simple analysis about the generation of the steady current for the case of $V_{j} =0 $ and $ F_{j} \neq 0 $ (we still set $ U=0 $). Here we assume that the steady state is periodic in space with the period of a unit cell. Thus, we focus on only a single cell which contains four sites. We have confirmed that the resulting final state of the numerical time evolution for a system with multiple cells under the periodic boundary condition actually satisfies this assumption. Further, in the present example where the pump and the loss in $F_j$ are balanced by themselves without introducing the imaginary $U$, we have found that the final steady state does not even show the overall phase rotation (i.e., the eigenvalue of the time-independent gDNLSE is zero) \cite{note:zeroenergy}. Therefore, for such a steady state, Eq.~(\ref{eq:2}) reads
\begin{equation}
\begin{aligned}
i\hbar \dfrac{d\psi_{j}}{dt} = -J (\psi_{j+1} + \psi_{j-1}) + i F_{j}\psi_{j}=0.
\end{aligned}
\label{eq:5} 
\end{equation}  
Then we can get the following relation,
\begin{equation}
\begin{aligned}
\psi_{j}= \frac{J}{i F_{j} } (\psi_{j+1} + \psi_{j-1}),
\end{aligned} 
\label{eq:6} 
\end{equation} 
for $j=1\sim 4$, with the following periodic boundary conditions: $\psi_{0}=\psi_{4}$ and $\psi_{5}=\psi_{1}$. Substituting Eq.~(\ref{eq:6}) for $j=1$ and $3$ into Eq.~(\ref{eq:3}), one can obtain the steady current $ \mathcal{J} $ given by
\begin{equation}
\begin{aligned}
\mathcal{J} = \frac{J^2}{\hbar} \left( \frac{1}{F_{1}} - \frac{1}{F_{3}} \right) (|\psi_{2}|^2 -|\psi_{4}|^2).
\end{aligned}  
\label{eq:7} 
\end{equation}  
Similarly, by substituting Eq.~(\ref{eq:6}) for $j=2$ and $4$ into Eq.~(\ref{eq:3}), one can obtain another expression of $\mathcal{J}$:
\begin{equation}
\begin{aligned}
\mathcal{J} = \frac{J^2}{\hbar} \left( \frac{1}{F_{2}} - \frac{1}{F_{4}} \right) (|\psi_{3}|^2 -|\psi_{1}|^2).
\end{aligned}  
\label{eq:8} 
\end{equation}  
Since the number of sites per cell (four sites) is even, it is impossible to be $\mathbb{P}$ symmetric around the middle of neighboring sites. Namely, the system can be $\mathbb{P}$ symmetric only around either of sites. When $ F_{j} $ is $\mathbb{P}$ symmetric around site $ j=1 $ or $ 3 $, $ F_{j}$ satisfies $ F_{2} = F_{4} $. Therefore, the current $ \mathcal{J} =0 $ from Eq.~(\ref{eq:8}). On the other hand, when $ F_{j} $ is $\mathbb{P}$ symmetric around $ j=2 $ or $ 4 $, it satisfies $ F_{1} = F_{3} $, so that $\mathcal{J} =0$ according to Eq.~(\ref{eq:7}). Thus the steady-state current is zero for a parity-inversion-symmetric $ F_{j} $. For $ F_{j} $ without the $\mathbb{P}$ symmetry, we have $ F_{1} \ne  F_{3} $ and $ F_{2} \ne  F_{4} $. Therefore, whether the current $ \mathcal{J} $ is zero or nonzero depends on the density distribution $|\psi_{j}|^2$ of the resulting steady state. Furthermore, the direction of the current also depends on $|\psi_{j}|^2$. Thus, we cannot tell the direction of the steady current before solving the steady state.

\begin{figure}[hbtp!]
	\centering	
	\includegraphics[width=0.98\linewidth]{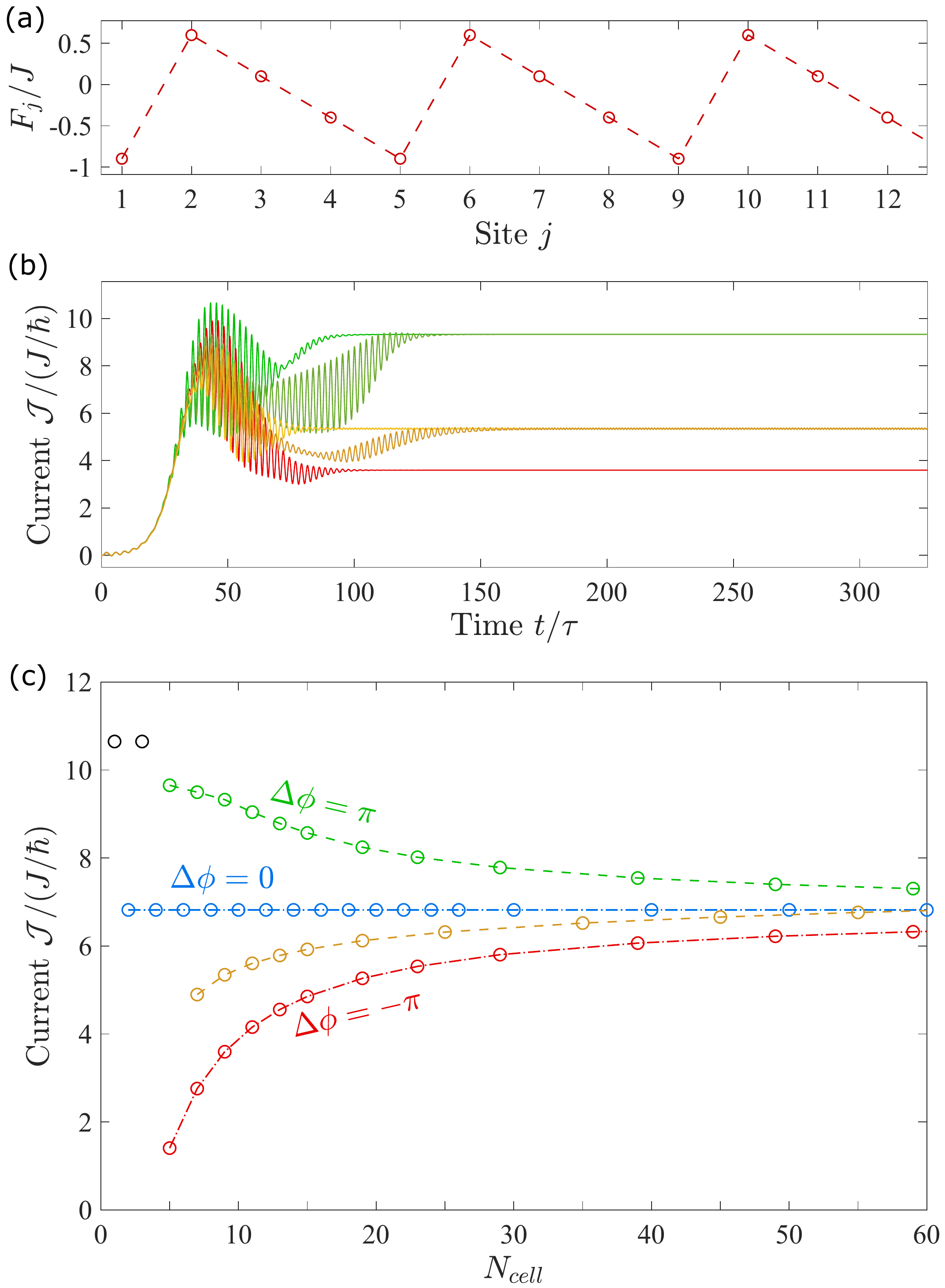} 
	\caption{Current and its steady value in the case of the nonzero imaginary part of $U$. (a) Ratchet potential form of the effective pump given by $F_{j} = F_{0}\times[1.2-\text{mod}(j+2, 4)]$, with the amplitude factor $F_{0}/J=0.5$. (b) Current as a function of time for several realizations with different values of the steady current under a given system size ($N_{\text{cell}}=9$). We initially prepare the state given by Eq.~(\ref{eq:4}) with random perturbation ($\xi_{j}\in [-0.05, 0.05]$). Examples of a few realizations of the time evolution of the system are shown in different colors. (c) Steady current as a function of the system size $N_{\text{cell}}$ for several series of $\Delta \phi$, each of which are connected by a dotted line with a different color. The current of the series of $\Delta \phi \ne 0$ converges to that of $\Delta \phi = 0$ in the limit of $N_{\text{cell}} \to \infty$. Here we set $U/J=-0.01i$.}
	\label{fig:2}	
\end{figure} 

Now, let us turn to discuss the effects of the nonlinear interaction. We first focus on the nonlinearity due to the imaginary part of $U$ and set the real part of $U$ to zero. Note that, due to the particle-particle interaction, the eigenvalue of the steady state is no longer zero in general, so that the above analysis is not applicable. Here, we consider the system with $N_{\text{cell}}$ cells (each cell still contains four sites) under the periodic boundary condition.
We set $V_j=0$ and take $F_j$ in the ratchet-potential form given by Eq.~(\ref{eq:fjasym}) with $F_0 = J/2$ [see Fig.~\ref{fig:2}(a)]. Now, we set the tuning offset $a = 1.2$ so that the pump and the loss in $F_j$ are not balanced by themselves unlike the previous case with $a=1$.
We initially prepare the state $\psi_{j}^{(0)}$ given by Eq.~(\ref{eq:4}) with the uniformly distributed random perturbation $\xi_{j} \in [-0.05, 0.05]$. As mentioned before, the imaginary part of $U$ is a gain-saturation coefficient, which prevents the magnitude of the wave function $\psi_{j}$ from diverging and stabilizes it to a finite value. On the other hand, nonlinearity due to the nonzero $U$ leads to the emergence of multiple steady states. Figure~\ref{fig:2}(b) shows the resulting time evolution of the current for several realizations, each of which is shown in a different color. Since there exist multiple steady states due to the nonlinearity caused by the nonzero imaginary part of $U$, realizations with different initial perturbations can end up with different values of the steady current.


\begin{figure}[t]
	\centering	
	\includegraphics[width=0.9\linewidth]{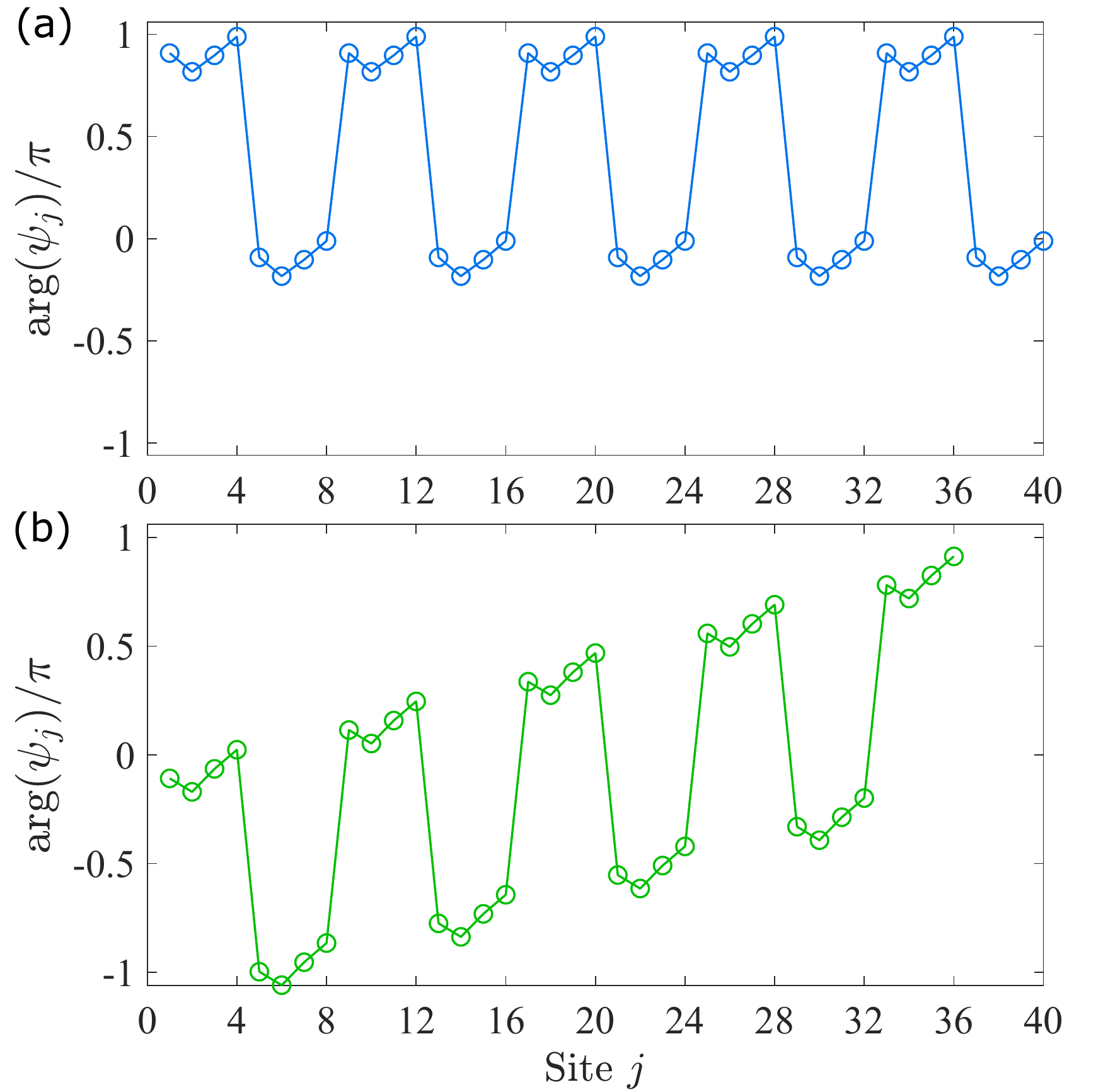} 
	\caption{Snapshots of the phase profile $\arg(\psi_j)$ of a steady state which belongs to the series of steady states characterized by the overall phase changes $\Delta\phi=0$ (a) and $\Delta\phi=\pi$ (b). Specifically, panel (a) is for the case of $N_{\text{cell}}=10$ at $t=300\tau$ for one of the realizations, and panel (b) is for the case of $N_{\text{cell}}=9$ at $t=300\tau$ for one of the realizations. The other parameters are the same as those in Fig.~\ref{fig:2}.}
	\label{fig:phase}	
\end{figure} 


Next, we discuss the system-size dependence of the steady current in this case. By looking into the resulting steady states, we find their wave function has the following two properties except for the cases of small system size with $N_{\text{cell}}=1$ and $3$: (i) the magnitude is periodic in space with period $d$ and (ii) the phase has an alternating $\pm \pi$ jump between each two adjacent cells on top of a global and monotonic phase change with an almost constant slope [e.g., the phase profile shown in Fig.~\ref{fig:phase}(b)]. Figure~\ref{fig:2}(c) shows the current $\mathcal{J}$ of the final steady state as a function of $N_{\text{cell}}$. At each $N_{\text{cell}}$, there are a number of possible values of the steady current $\mathcal{J}$, each of which corresponds to a different steady state. Each line in this figure shows a series of the steady current characterized by an overall phase change $\Delta\phi$ throughout the system due to the component of the constant slope (i.e., the phase change except for the contribution by the $\pm\pi$ jumps), which is an integer multiple of $\pi$. For instance, Figs.~\ref{fig:phase}(a) and \ref{fig:phase}(b) show the phase profile of a steady state which belongs to the series characterized by the overall phase changes $\Delta\phi=0$ and $\pi$, respectively. It is noted that, although there are many possible values of the steady current with different values of $\Delta\phi$ at each $N_{\text{cell}}$, the states in the series of larger $|\Delta\phi|$ are hard to realize and practically irrelevant since they are energetically unfavorable. Figure~\ref{fig:2}(c) tells that the steady currents of all the relevant series with small values of $|\Delta\phi|$ converge to a single value given by the series of $\Delta \phi=0$ in the limit of $N_{\text{cell}} \to \infty$. In addition, since the steady current of the series of $\Delta \phi=0$ is constant with respect to $N_{\text{cell}}$, the steady current in the infinite system can be evaluated by the one for $\Delta \phi=0$ at any even $N_{\text{cell}}$ (the series of $\Delta\phi=0$ does not exist for odd $N_{\text{cell}}$ as we shall explain below). 

The constancy of $\mathcal{J}$ of the series of $\Delta\phi=0$ can readily be explained from the abovementioned property (ii) on the phase of the resulting steady states. Since the overall constant phase gradient of the steady states in the series of $\Delta\phi=0$ is zero by definition, the change of the phase is only by the alternating $\pm \pi$ phase jumps between adjacent cells [see, e.g., Fig.~\ref{fig:phase}(a)]. Therefore, $N_{\text{cell}}$ must be even because the overall phase change has to be an integer multiple of $2\pi$ in order to satisfy the single valuedness of the wave function under the periodic boundary condition. Since there is no overall constant phase gradient for the $\Delta\phi=0$ series, when adding two cells to the system, the wave function just repeats one more period, and consequently the average current per cell $\mathcal{J}$ is unchanged.

The $N_{\text{cell}}$ dependence of $\mathcal{J}$ for the other series of $\Delta\phi \ne 0$ can also be explained as follows. First, we note that, for the series of $\Delta\phi$ equal to an even (odd) multiple of $\pi$, $N_{\text{cell}}$ must be even (odd) to satisfy the periodic boundary condition. Due to the nonzero overall phase gradient of the steady states for the series of $\Delta\phi \ne 0$, the wave function deviates from that of the series of $\Delta \phi=0$, so does the average current. When $N_{\text{cell}}$ is reduced with $\Delta\phi$ fixed at a nonzero value, this deviation due to the nonzero overall phase gradient becomes more noticeable for smaller system size since the phase gradient increases as $\sim 1/N_{\text{cell}}$. Conversely, when $N_{\text{cell}}$ is increased, the overall phase gradient gets smaller and smaller so that the steady state wave function becomes more similar to the one of $\Delta\phi = 0$ and the current converges to that of $\Delta \phi=0$ in the limit of $N_{\text{cell}} \to \infty$.

In addition to the series of steady states, the long-time averaged current for a series of ``quasisteady states'' is also shown in Fig.~\ref{fig:2}(c) (the brown dashed curve). These quasisteady states, which emerge due to the nonlinearity, are not the true steady states, but are oscillating in time. Indeed, their current shown by the brown lines in Fig.~\ref{fig:2}(b) oscillates in time even at large $t$ (its amplitude is too small to clearly be discerned in this figure). However, although the current actually oscillates even in the limit of large time $t$, the oscillation is regular and the oscillation amplitude becomes negligibly small for sufficiently large $N_{\text{cell}}$. It is noted that, even for the series of the quasisteady states, its long-time averaged current also asymptotically converges to the series of $\Delta\phi = 0$ as shown in Fig.~\ref{fig:2}(c) \cite{note:asymptotic}.

Let us remark on the two isolated points at $N_{\text{cell}}=1$ and $3$ in Fig.~\ref{fig:2}(c) which do not belong to any series of steady states.
For steady states of these two isolated points, we cannot distinguish between the $\pm \pi$ phase jumps and the overall constant phase gradient due to their short spatial period. Furthermore, their eigenvalue is zero at any value of the amplitude $F_0$ of the imaginary ratchet potential (on the other hand, the other steady states of any series characterized by $\Delta \phi$ have a nonzero real eigenvalue and its value changes by $F_0$). In addition, we find that, for sufficiently large amplitude $F_0$ of the ratchet potential given by Eq.~(\ref{eq:fjasym}), there exists a unique steady state for all $N_{\text{cell}}$ (i.e., the same unique steady state for any $N_{\text{cell}}$). It is noted that this unique steady state for large $F_0$ is adiabatically connected to the steady state of the abovementioned two isolated points, which appears only at small $N_{\text{cell}}$ when $F_0$ is small. By directly looking into the phase of the unique steady state numerically obtained for large $F_0$, we find that the phase difference between every pair of adjacent sites is either $\pi/2$ or $-\pi/2$.
This can be understood as follows. For the steady state with a zero eigenvalue, the equation of motion (\ref{eq:2}) reads
\begin{align}
0 = -J (\psi_{j+1} + \psi_{j-1}) +U|\psi_{j}|^2\psi_{j}  + iF_{j}\psi_{j}.
\label{eq:8-2} 
\end{align}
This equation can be rewritten as
\begin{equation}
\begin{aligned}
0 = -J \left(\dfrac{|\psi_{j+1}|}{|\psi_{j}|} e^{i\theta_{j+1,\, j}} + \dfrac{|\psi_{j-1}|}{|\psi_{j}|}e^{i\theta_{j-1,\, j}} \right) +U|\psi_{j}|^2 + iF_{j},
\end{aligned}
\label{eq:8-3} 
\end{equation} 
where $\theta_{j\pm 1,\, j} \equiv \arg{(\psi_{j\pm 1})} - \arg{(\psi_j)}$ is the phase difference between $\psi_{j\pm1}$ and $\psi_{j}$. Note that $F_j$ here spatially oscillates between positive and negative values since we consider a system where the pump and the loss can be balanced. If $F_j$ at site $j$ becomes a large negative value (i.e., the loss rate at site $j$ becomes large), the number of particles $|\psi_j|^2$ at this site gets small, so that the interaction term $U|\psi_j|^2$ becomes negligible. In addition, in the case of our imaginary potential, if it is negative at site $j$, it takes a positive value at one of the neighboring sites, $j-1$ or $j+1$ [see Fig.~\ref{fig:2}(a)], where the number of particles $|\psi_{j-1}|^2$ or $|\psi_{j+1}|^2$ is much larger than at site $j$ and the other neighboring site (i.e., for site $j=1$ with $F_j<0$, its neighboring site $j+1=2$ has the positive imaginary potential $F_{j+1}>0$; for site $j=4$ with $F_j<0$, site $j-1=3$ has $F_{j-1}>0$). Therefore, the real part of Eq.~(\ref{eq:8-3}) for $j=1$ and $4$ approximately reduces to (the upper sign for $j=1$ and the lower one for $j=4$)
\begin{align}
  0 \simeq -J \frac{|\psi_{j\pm 1}|}{|\psi_j|} \cos{\theta_{j\pm 1,\, j}}\,,\label{eq:realpartgpe}
\end{align}
which can be satisfied if and only if $\theta_{2,\, 1} \simeq \pm\pi/2$ and $\theta_{3,\, 4} \simeq \pm \pi/2$, respectively, since $|\psi_{j \pm 1}|$ is nonzero. On the other hand, for sites with a positive imaginary potential (i.e., sites $j=2$ and $3$), one of the neighboring sites has the negative imaginary potential (i.e., site $j-1=1$ for site $j=2$, and site $j+1=4$ for site $j=3$), where the number of particles becomes negligible for sufficiently large $F_0$. Therefore, the real part of Eq.~(\ref{eq:8-3}) for sites $j=2$ (the upper sign) and $3$ (the lower sign) approximately reads the same equation as Eq.~(\ref{eq:realpartgpe}), and thus we get $\theta_{3,\, 2} \simeq \pi/2$ or $-\pi/2$. Since $\theta_{2,\, 1}$, $\theta_{3,\, 2}$, and $\theta_{4,\, 3}$ are $\pi/2$ or $-\pi/2$, the remaining phase difference $\theta_{1,\, 4}$ between sites $1$ and $4$ should also be $\pi/2$ or $-\pi/2$ due to the periodicity of the wave function. Regarding the magnitude of the wave function at sites with positive $F_j$ (i.e., $j=2$ and $3$), we get $|\psi_j| \simeq \sqrt{F_j/iU}$ since the number of particles $|\psi_j|^2$ at these sites is large and the imaginary part of Eq.~(\ref{eq:8-3}) approximately reads
\begin{align}
  0 \simeq i^{-1} U|\psi_j|^2 + F_j\,.
\end{align}
Lastly, the magnitude $|\psi_j|$ at sites with negative $F_j$ (i.e., $j=1$ and $4$) is $|\psi_j| \simeq 0$ for sufficiently large $F_0$. Here, we have seen that the phase structure and the amplitude profile of the wave function are uniquely determined: this concludes there is only one final steady current for the imaginary ratchet potential in the form of Fig.~\ref{fig:2}(a) with sufficiently large $F_{0}$.

Finally, we briefly discuss the effect of the nonzero real part of $U$. The qualitative results discussed above do not change when $\Re(U)$ is smaller than or comparable to $\Im(U)$: There are still a number of series of the steady current characterized by $\Delta\phi$ of an integer multiple of $\pi$. The steady current of the $\Delta\phi = 0$ series is constant with respect to $N_{\text{cell}}$, and the current of the other series converges to the value of the $\Delta\phi = 0$ series in the thermodynamic limit. An important benefit of introducing the nonzero real part of $U$ is that it can stabilize the aforementioned quasisteady states when $\Re(U)$ is comparable with $\Im(U)$.

\section{Continuous model}\label{sec:continuous model}

Next, we are going to verify the results obtained by the simple discrete model in Sec.~\ref{sec:discrete model} using a more realistic continuous model, a generalized Gross-Pitaevskii equation (gGPE). The gGPE, which can describe exciton-polariton condensates in a semiconductor microcavity \cite{Keeling2008PRL, Eastham2008PRB, Wouters2010PRL, Keeling2011Contempphys, Moxley2016PRA}, is given by 
\begin{equation}
\begin{aligned}
i\hbar\frac{\partial \Psi(x,t)}{\partial t} =\Big[  &-\frac{\hbar^2}{2m} \frac{\partial^2}{\partial x^2} +g|\Psi(x,t)|^2 + V(x) \\
                                               &+\frac{i\hbar}{2}\left(P(x) - \gamma(x) - \eta|\Psi(x,t)|^2\right) \Big] \Psi(x,t),
\end{aligned}
\label{eq:9}
\end{equation}
where $\Psi(x,t)$ is the condensate wave function of polaritons, $m$ is the effective mass of the polariton, $g$ is the polariton-polariton interaction strength in the condensate, $V$ is the external potential, $P$ is the pumping rate of the polariton condensate determined by the power of the pumping laser and the density of the reservoir excitons, and $\gamma$ is the loss rate of the polaritons due to the leakage of photons from the microcavity. Note that the effect of the presence of reservoir excitons on the polariton condensate is phenomenologically captured by the $\eta$ term which introduces the gain-saturation effect of the polariton condensate \cite{Keeling2008PRL, Keeling2011Contempphys}. This gain-saturation term corresponds to the imaginary part of $U$ in the discrete model. We define $ F \equiv P - \gamma $ as the effective pump which can be regarded as a single external imaginary potential as a whole. As in the previous section, we consider $P(x)$ and $\gamma(x)$ are periodic in space with period $d$ of a unit cell, and so is $F(x)$. Here as well, we consider the system with $N_{\text{cell}}$ cells under the periodic boundary condition. The average current of the polariton condensate is defined as 
\begin{align}
\mathcal{J} = \frac{i\hbar}{2m N_{\text{cell}} d} \int \left(\Psi \frac{\partial \Psi^{*}}{\partial x} - \Psi^{*} \frac{\partial \Psi}{\partial x} \right) dx.
\label{eq:10}
\end{align}
In the calculation of the gGPE in this section, we take $ E_{0}\equiv\hbar^{2}\pi^{2}/(2md^{2})$ as the unit of energy and $\tau \equiv \hbar/E_{0} $ as the unit of time.

\begin{figure}[t!] 
	\centering	
	\includegraphics[width=0.98\linewidth]{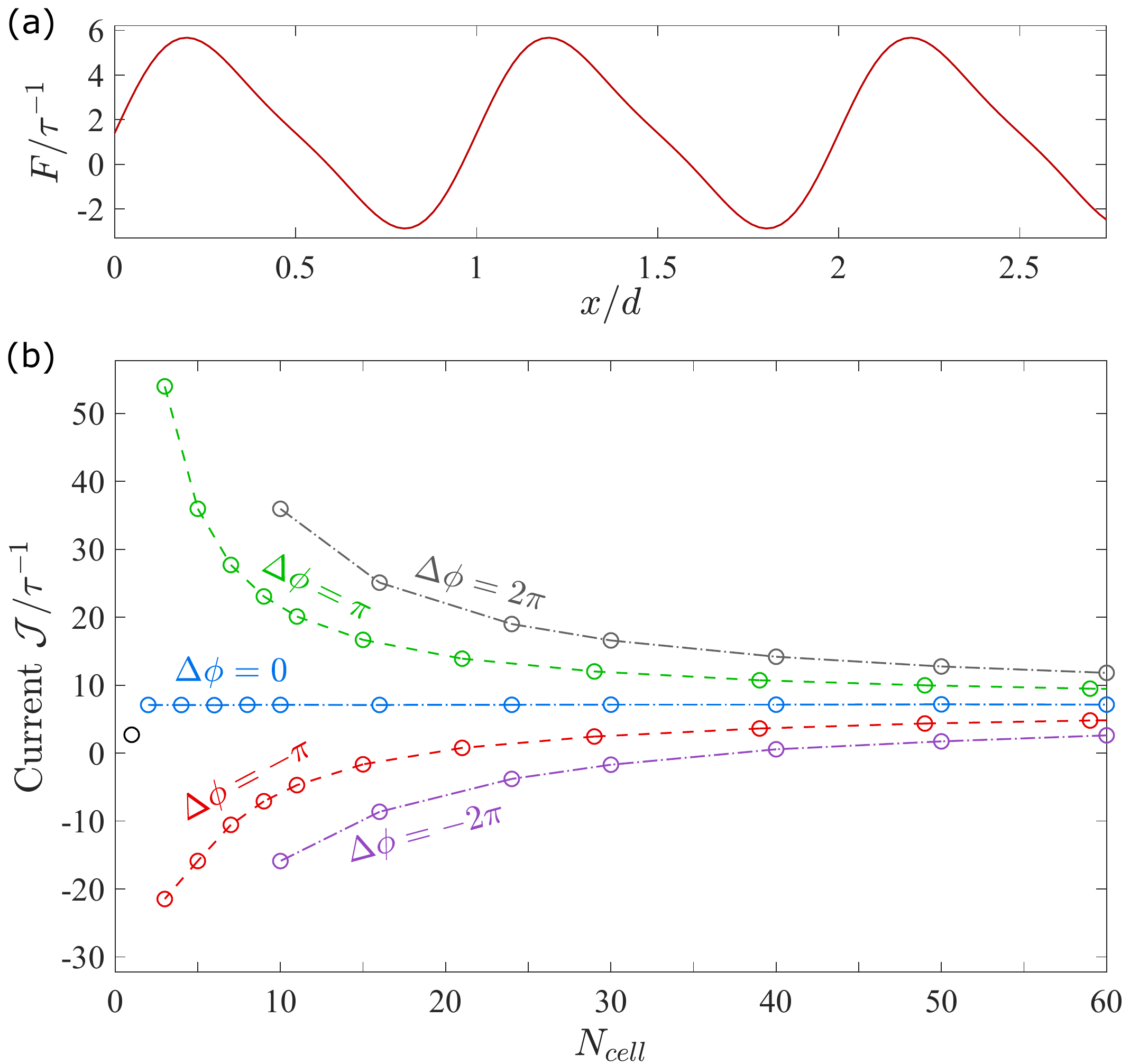} 
	\caption{Ratchet-potential form of the effective pump $F(x)$ and the resulting persistent current with gain-saturation ($\eta \ne 0$). (a) Effective pump as a function of $x$ given by $F(x)/\tau^{-1}= 4[\sin(2\pi x/d)+ 0.2 \sin(4\pi x/d)+0.35]$. (b) Steady current as a function of the system size $N_{\text{cell}}$. The series of $\Delta \phi= 0$, $\pm \pi$, $\pm2 \pi$ are shown, and each series of the data points is connected by a dotted line with a different color. The steady states with larger $|\Delta\phi|$ are not obtained at small $N_{\text{cell}}$ in our simulations since they are energetically unfavorable. Here we set $V=g=0$ and $\eta /\tau^{-1}=0.05$.}
	\label{fig:3}
\end{figure}

The effect of the violation of the $\mathbb{P}$ symmetry of the real and the imaginary potentials on the current is qualitatively the same as in the case of the discrete model discussed in Sec.~\ref{sec:discrete model}. The imaginary potential $F(x)$ without the $\mathbb{P}$ symmetry can generate a net current. Here we consider an imaginary ratchet potential in the following form:
\begin{align}
F(x)= F_{0}[\sin(kx)+ \alpha \sin(2kx)+\beta],
\label{eq:11}
\end{align}
where $F_{0}$ is the amplitude of the effective pump, $\alpha$ is the shape coefficient of the ratchet, $\beta$ is the offset to tune the gain and the loss, and $k$ is set by the spatial period $d$ of the system as $k=2\pi/d$. Here, we take $F_{0}=4 E_0/\hbar = 4\tau^{-1}$, $\alpha=0.2$, and $\beta=0.35$. The shape of the imaginary ratchet potential $F(x)$ is shown in Fig.~\ref{fig:3}(a). For simplicity, we neglect the interaction term and the external potential, i.e., $g=V=0$. We take the following initial state whose particle number per cell is normalized to unity: 
\begin{align}
\Psi^{(0)}(x)=\frac{\displaystyle{ \sqrt{N_{\text{cell}}} \left[1+\xi(x) \right] } }    { \sqrt{  \displaystyle{ \int \left| 1+\xi(x) \right|^{2} dx}} },
\label{eq:12}
\end{align}
where $\xi(x)$ is the random perturbation. For each realization of the time evolution, we take the random perturbation $\xi(x) \in [-0.05, 0.05]$ with a uniform distribution at each grid point separated by $\Delta x=0.02d$ \cite{note:Delta x} in our simulation.

Simulation results of the continuous model qualitatively agree with those of the discrete model shown in Sec.~\ref{sec:discrete model}. This validates the main results obtained by the discrete model. Resulting from the nonlinearity of the gain-saturation term ($\eta$ term), there appear multiple final steady states at each $N_{\text{cell}}$ as we can see in Fig.~\ref{fig:3}(b). As in the case of the discrete model, each steady state can be characterized by the overall phase change $\Delta\phi$, and here we show the series of $\Delta \phi=0$, $\pm \pi$, and $\pm2 \pi$. The series of $\Delta\phi=0$, which exists at even $N_{\text{cell}}$, shows a constant value of the steady current, and the steady current of the other series of $\Delta\phi \ne 0$ converges to that of $\Delta\phi=0$ in the limit of $N_{\text{cell}} \to \infty$. It is also noted that there is an isolated point at $N_{\rm cell}=1$ shown by a black circle in Fig.~\ref{fig:3}(b). This isolated point is essentially the same as the aforementioned two isolated points in Fig.~\ref{fig:2}(c) obtained in the discrete model. For sufficiently large amplitude $F_0$ of the ratchet potential given by Eq.~(\ref{eq:11}), there exists a unique steady state in common for all $N_{\text{cell}}$, and this unique steady state is adiabatically connected to the steady state of this isolated point.

\section{Demonstration by the coupled Gross-Pitaevskii equation}\label{sec:coupled Gross-Pitaevskii equation}

In this section, we give a demonstration of generating a persistent current of exciton-polariton condensates in a 2D semiconductor microcavity with a periodic loss rate with period $d$ in one direction, which we take to be the $x$ direction. Although the system is 2D, the structure is 1D along the $x$ direction and is uniform in the transverse ($y$) direction. Assuming the transverse component of the polariton wave function $\Psi_{\rm 2D}(x,y,t)$ is homogeneous, we have $\Psi_{\rm 2D}(x,y,t)= \Psi(x,t)/\sqrt{L_y}$, where $L_y$ is the transverse length of the system. Thus, we can reduce the 2D open-dissipative GP equation, a standard model widely used to describe the exciton-polaritons \cite{Wouters2007PRL, Carusotto2013RMP, Ma2015PRB, Chestnov2016PRB}, into a 1D problem:   
\begin{equation}
\begin{aligned}
i\hbar \frac{\partial \Psi(x,t)}{\partial t} =\Big[&-\frac{\hbar^2}{2m} \frac{\partial^2}{\partial x^2} + V +g_{c}|\Psi(x,t)|^2\\
&+ \frac{i\hbar}{2}(nR - \gamma_{c}) +ng_{R} \Big] \Psi(x,t),
\end{aligned}
\label{eq:13}
\end{equation}
coupled with the following rate equation describing the population dynamics of the reservoir excitons:
\begin{align}
\frac{\partial n}{\partial t} &= P-n\gamma_{R}-nR|\Psi(x,t)|^2,
\label{eq:14}
\end{align}
where $n$ is the 1D density of the reservoir excitons; $P$ is the gain rate of the reservoir excitons from the laser pump; $g_{c}$ and $g_{R}$ are the two-body interaction strengths of the polaritons in the condensate and the reservoir excitons, respectively; $R$ is the scattering rate between the reservoir excitons and the condensate polaritons; and $\gamma_{c}$ and $\gamma_{R}$ are the loss rates of the condensate polaritons and the reservoir excitons, respectively. Note that $g_{c}$, $g_{R}$, $R$, and $P$ in Eqs.~(\ref{eq:13}) and (\ref{eq:14}) are given by $g_c \equiv g^{\rm 2D}_{c}/L_y$, $g_R \equiv g^{\rm 2D}_{R}/L_y$, $R \equiv R^{\rm 2D}/L_y$, and $P \equiv P^{\rm 2D}/L_y$, where the quantities with the superscript ``2D'' represent the corresponding quantities in the original 2D model. Since the wave function in the transverse direction is assumed to be the homogeneous ground state, the results of the flux (and intensive properties of the system as well) do not depend on the system size $L_y$ in the transverse direction. Therefore, for convenience, we set $L_y$ to be equal to $d$ in our simulations.

\begin{figure}[tb!] 
	\centering	
	\includegraphics[width=0.99\linewidth]{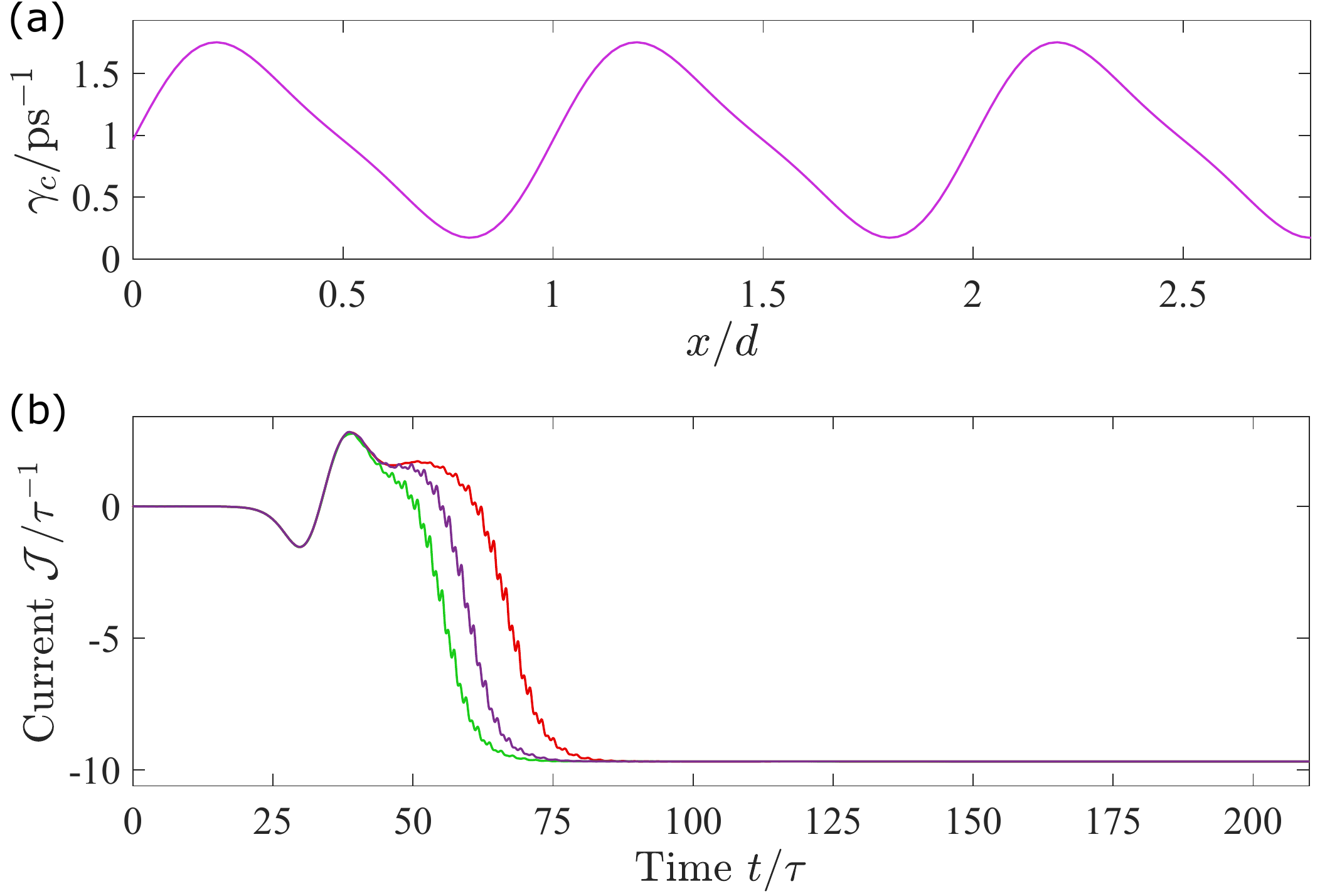} 
	\caption{Ratchet-potential form of the loss rate $\gamma_{c}(x)$ and the resulting persistent current of the exciton-polariton condensate in a semiconductor microcavity. (a) The loss-rate profile $\gamma_{c}(x)=0.6[\sin(2\pi x/d)+ 0.2\sin(4\pi x/d)+1.3]$\, ps$^{-1}$. In the simulation, we set $d=2.8$~$\mu$m. (b) Current as a function of time for a few realizations (shown in different colors) with the same value of the steady current. Here we set $N_{\text{cell}}=10$. The initial state is given by Eq.~(\ref{eq:12}) with random perturbation $(\xi_{j}\in [-0.05, 0.05])$.}
	\label{fig:4}
\end{figure}

We use the parameter values from the experiment by Roumpos {\it et al.} \cite{Roumpos2011natphys}: $g^{\rm 2D}_{c} =6 \times 10^{-3}\, \text{meV}\, \mu \text{m}^2$, $ g^{\rm 2D}_{R} =2 g^{\rm 2D}_{c} $, and $ R^{\rm 2D} =0.01\, \text{ps}^{-1}\, \mu \text{m}^2 $. Besides, we take the pumping strength $P^{\rm 2D}=50\,\text{ps}^{-1}\, \mu \text{m}^{-2} $, which is also comparable to the value in the above experiment. Here, we consider the loss rate $\gamma_{c}(x)$ of the polariton condensate to have a ratchet-potential shape (i.e., an imaginary ratchet potential) with period $d$ of a unit cell as shown in Fig.~\ref{fig:4}(a), and the loss rate $\gamma_R$ of reservoir excitons is given by $\gamma_{R}(x) = 1.5\gamma_{c}(x)$. One can obtain the ratchet-potential form of $\gamma_{c}(x)$ and $\gamma_{R}(x)$ by tuning the quality factor of the microcavity by etching the distributed Bragg mirrors \cite{Dousse2008PRL, Tanese2013natcommun, Jayaprakash2020ACSPhoto}. Such a ratchet-potential form of $\gamma_{c}(x)$ and $\gamma_{R}(x)$ breaks the $\mathbb{P}$ symmetry and the $\mathbb{T}$ symmetry of the Hamiltonian simultaneously, so that a persistent current can be generated.

In the simulation, we consider the system under the periodic boundary condition. To show the steady current in the thermodynamic limit, we show the results whose steady state is the one with $\Delta\phi=0$ in the following. Thus, while we set $N_{\text{cell}}=10$ in the particular simulation shown below, the resulting steady current is independent of the choice of $N_{\text{cell}}$. We take $ E_{0}=\hbar^{2}\pi^{2}/(2md^{2}) $ as the unit of energy and $\tau= \hbar/E_{0} $ as the unit of time. For the effective mass of the polariton $ m\approx9\times10^{-5}m_\text{e} $ and the lattice constant $ d=2.8\, \mu \text{m} $ as in the experiment \cite{Lai2007nat}, we get $ E_{0} \approx 0.533$~meV and $ \tau \approx 1.24$~ps. We prepare the initial state in the form of Eq.~(\ref{eq:12}) with uniformly distributed random perturbation $\xi (x) \in [-0.05, 0.05]$ at each grid point whose interval is $\Delta x = 0.02d$. The average current during the time evolution of the system is shown in Fig.~\ref{fig:4}(b), which shows a stable current of $\approx -9.67\tau^{-1} (=-7.80~\text{ps}^{-1})$ after $t \gtrsim 85 \tau$. This amount of the current can be compared with the critical current. The critical velocity of the polariton condensate can be estimated by the sound velocity $c_s = \sqrt{g_c n_c/m}$ of the corresponding homogeneous system with the same average density \cite{Wouters2010PRL}, where $n_c$ is the average density of the polariton condensate. Since $n_c \approx 64.6 d^{-1}$ in the resulting steady state of the present example, the critical current is estimated as $\sim n_c c_s \approx 33.6$~ps$^{-1}$. Therefore, the obtained steady current amounts to $\sim 23\%$ of the critical current.

\section{Summary and conclusion}
We have proposed a scheme to generate a persistent current in driven-dissipative nonequilibrium systems. The time-reversal ($\mathbb{T}$) symmetry is intrinsically broken due to the dissipation (loss) in these systems. We consider the static, position-dependent loss rate in the ratchet-potential form to break the parity-inversion ($\mathbb{P}$) symmetry of the systems. Such a loss-rate profile which simultaneously breaks the $\mathbb{T}$ symmetry and the $\mathbb{P}$ symmetry of the Hamiltonian can generate a net current. Our scheme neither requires any dynamic drive nor nonlocal dissipation, and it can generate a nonzero current even in the thermodynamic limit. Furthermore, by introducing a quantum wire structure, the minimum width of the flow can be reduced to the nanometer scale, which is crucial to make a high-density integrated circuit. These points are advantages of our scheme over the other existing ones.

First, we have discussed the basic physics of our scheme using the simple discrete driven-dissipative Gross-Pitaevskii model. We have found that the gain-saturation term can stabilize the system into a steady state. Although multiple steady states with different amounts of current emerge due to the nonlinearity of the gain-saturation term, their current asymptotically approaches the same value in the limit of infinite system size. Qualitative results obtained in the discrete model have been verified by the more realistic continuous model. Finally, we have demonstrated the experimental feasibility of our scheme to generate a persistent current in exciton-polariton condensates in a semiconductor microcavity. 

For future prospects, it would be interesting to manipulate the polariton current in a semiconductor microcavity by designing a nontrivial 2D pattern of the imaginary potential. Quantum interferometers \cite{Sturm2014ncomms} and ``polariton circuits'' \cite{Liew2008PRL, Liew2010PRB, Xu2020PRAppl, Mirek2021nanoLett} can be implemented in this 2D platform. In addition, degrees of freedom of the exciton-spin can be taken into consideration \cite{Shelykh2005pssb, Leyder2007natphys, Larionov2010PRL, Askitopoulos2015PRB, Askitopoulos2016PRB, Xu2017PRB, Klaas2019PRB}. The transport of spinor exciton-polariton condensates \cite{Leyder2007natphys, Pinsker2014PRL, Caputo2019CommunPhys} might also be an interesting problem with rich applications.

\begin{acknowledgments}
This work was supported by the NSF of China (Grants No.~11975199 and No.~11674283), by the Zhejiang Provincial Natural Science Foundation Key Project (Grant No.~LZ19A050001), and by the Zhejiang University 100 Plan.
\end{acknowledgments}


\begin{thebibliography}{99}
\bibitem{note:zeroenergy} When $U=0$ and $F_j \ne 0$, and the pump and loss are in balance to have a steady state, our discrete model with four sites per unit cell has only one steady state and its eigenvalue is always zero.
\bibitem{note:asymptotic} More precisely, due to the small oscillation of the quasisteady states, their long-time averaged current in the limit of $N_{\text{cell}} \rightarrow \infty$ slightly deviates from the steady current of the series of $\Delta\phi = 0$. Depending on the series of the quasisteady states, the asymptotic value in the limit of $N_{\text{cell}} \rightarrow \infty$ can be either larger or smaller than the steady current of $\Delta\phi = 0$ series.
\bibitem{note:Delta x} For small $N_{\text{cell}}\ (< 10)$, we take a smaller $\Delta x$. We have confirmed that all the results shown in this paper converge with respect to $\Delta x$.
\end{thebibliography}
\end{document}